\documentclass[aps,prd,superscriptaddress,floatfix,nofootinbib,notitlepage]{revtex4-1}

\usepackage{color,amsmath,amssymb,multirow,booktabs,graphicx,mathtools,feynmp} 

\makeatletter
\@ifundefined{pdfoutput}{}{\DeclareGraphicsRule{*}{mps}{*}{}}
\makeatother
\newcommand{\arxiv}[1]{\href{http://arxiv.org/abs/#1}{arXiv:#1}}

\newcommand\one{\leavevmode\hbox{\small1\normalsize\kern-.33em1}}

\newcommand{\lag}{\mathcal{L}}

\newcommand{\qqquad}{\qquad \qquad}
\newcommand{\qqqquad}{\qquad \qquad \qquad}

\newcommand{\met}{\slashchar{E}_T}

\newcommand{\gev}{\text{GeV}}

\def\slashchar#1{\setbox0=\hbox{$#1$}           % set a box for #1
   \dimen0=\wd0                                 % and get its size
   \setbox1=\hbox{/} \dimen1=\wd1               % get size of /
   \ifdim\dimen0>\dimen1                        % #1 is bigger
      \rlap{\hbox to \dimen0{\hfil/\hfil}}      % so center / in box
      #1                                        % and print #1
   \else                                        % / is bigger
      \rlap{\hbox to \dimen1{\hfil$#1$\hfil}}   % so center #1
      /                                         % and print /
   \fi}

\newcommand{\eg}{\textsl{e.g.}\;}

\newcommand{\be}{\begin{eqnarray*}}
\newcommand{\ee}{\end{eqnarray*}}

\newcommand{\bee}{\begin{eqnarray}}
\newcommand{\eee}{\end{eqnarray}}
\newcommand{\beeq}{\begin{equation}}
\newcommand{\eeeq}{\end{equation}}

\begin{document}

\title{Dark Matter from Electroweak Single Top Production}

\author{Tilman Plehn}
\affiliation{Institut f\"ur Theoretische Physik, Universit\"at Heidelberg, Germany}

\author{Jennifer Thompson}
\affiliation{Institut f\"ur Theoretische Physik, Universit\"at Heidelberg, Germany}

\author{Susanne Westhoff}
\affiliation{Institut f\"ur Theoretische Physik, Universit\"at Heidelberg, Germany}

\begin{abstract}
  Dark matter scenarios with spin-0 mediators in the $s$-channel have
  be tested in well-established processes with missing energy, such as
  top-pair- and mono-jet-associated production. We suggest electroweak
  single top production in association with a dark matter pair as an
  alternative channel. Based on a realistic analysis for the LHC at 13
  TeV, we demonstrate how to efficiently discriminate between the
  signal and Standard-Model background using event kinematics. With
  $300\,\text{fb}^{-1}\,(3\,\text{ab}^{-1})$ of data, on-shell scalar mediators with a coupling strength $g_S^t=1$ to top quarks can be probed up to masses of $180\,(360)~\gev$. Single-top-associated dark matter production should thus
  be included as an independent search channel in the LHC dark matter program.
  \end{abstract}

\maketitle
\tableofcontents
\begin{fmffile}{feynman}
\newpage

%%%%%%%%%%%%%%%%%%%%%%%%%%%%%%%%%%%%%%%%%%%%%%%%%%%%%%%%%%%%%%%%%%%%%%
\section{Dark matter coupling to top quarks}
Searches for particle dark matter with masses
below the TeV scale and sizable couplings to the Standard Model (SM) 
are at the heart of the LHC program~\cite{review}. While the invisible state has to be a new particle, the mediators responsible
for its interaction with the Standard Model can either be known
particles or new particles. To allow for a stable dark matter candidate, we assume an appropriate conserved parity among the new dark
particles. In renormalizable models with a single dark matter state,
the mediators have to be SM bosons not coupling to the electric or
color charges, i.e., the $Z$ or the Higgs boson. However, models of
thermal dark matter relying on these mediators to predict the observed
relic density are very strongly constrained by direct detection
experiments~\cite{Escudero:2016gzx}.

Similarly, in models where the interaction of a dark matter pair with
the SM is mediated by a new particle through an $s$-channel process,
such mediators have to be color and electrically neutral
bosons~\cite{simplified,Agrawal:2010fh}. New gauge bosons, for which
gauge invariance typically implies universal couplings to all three
fermion generations, have come under strong pressure by LHC
searches involving interactions with light quarks~\cite{lhc-searches}. Vector and axial-vector bosons are
therefore disfavored as mediators. Here we consider the hypothesis of
dark matter coupling to a scalar or pseudo-scalar mediator, like for
example a heavy Higgs boson in a two-Higgs doublet model. To avoid
large flavor-changing neutral currents, such a new scalar ought to
have flavor-hierarchical couplings to quarks, mimicking the Yukawa
couplings of the Higgs boson in the SM. Observables with top quarks
are thus expected to be the dominant signatures of scalar-mediated
dark matter production at
colliders~\cite{cheung_kentarou,chicago}. Signals of dark matter
production in association with a pair of top quarks have been studied
in detail~\cite{aachen_louvain,haisch_re,giacomo}. The impact of final states
with a single top quark on searches for top-pair-associated production
has only been noticed recently~\cite{zurich}. This raises the question
whether single-top-associated dark matter production should be
explored as a {\it signal} at the LHC.

In this study, we show that the discovery prospects for spin-0-mediated
dark matter production in association with a single top quark are
comparable with top-pair-associated production. We suggest a dedicated search strategy for the
process\footnote{This process is different from the established
  mono-top signal, which does not involve the additional jet and
  probes different scenarios of new
  physics~\cite{monotop,our_monotop,Goncalves:2017soe}.}
\begin{align}
pp \to t j \, \chi \bar \chi,
\end{align}
where $\chi$ denotes the dark matter candidate and $j$ refers to a
hard jet originating from electroweak single top production via the
$t$-channel partonic process $qb\to q't$. Rather than using this
channel to strengthen the existing top-pair analysis we propose to
perform a dedicated single top analysis. If we can separate
the two signal phase space regions, a combined analysis will be
possible because the underlying dark matter hypothesis in both cases
is identical.

%%%%%%%%%%%%%%%%%%%%%%%%%%%%%%%%%%%%%%%%%%%%%%%%%%%%%%%%%%%%%%%%%%%%%%
\subsection{Fermion dark matter with spin-0 mediator}
\label{sec:model}

The simple model we consider consists of two new states in the dark
sector, a Dirac fermion dark matter candidate $\chi$ and a heavy
scalar ($S$) or pseudo-scalar ($P$) mediator. Our scenario is similar
to a subset of electroweakinos and new Higgs bosons in the MSSM,
combined with a switch from Majorana fermion to Dirac fermion. A
renormalizable mediator coupling to SM quarks requires the mediator to
transform non-trivially under the electroweak gauge group. This
generally implies Higgs-portal and potentially new gauge couplings,
which we neglect because they are model-dependent. Here we assume that additional particles in the dark sector have
no significant impact on our observables. The mediator couplings to SM
fermions are assumed to be flavor-hierarchical, i.e., proportional to
the SM Yukawa couplings. The leading interactions of the new particles
are then described by either of the Lagrangians
\begin{align}
\lag_S &\supset 
  g_S^{\chi} \, (\bar{\chi}\chi) \; S 
+ g_S^t \frac{m_t}{v} \,(\bar{t}t) \; S, \notag \\
\lag_P &\supset 
  i g_P^{\chi} \, (\bar{\chi}\gamma_5 \chi) \; P 
+ i g_P^t \frac{m_t}{v}\, (\bar{t} \gamma_5 t) \; P \; ,
\label{eq:lag}
\end{align}
where $v=246~\gev$ is the vacuum expectation value of the Higgs
field. The couplings $g_{S,P}^\chi$ and $g_{S,P}^t$ to
dark matter and top quarks, respectively, are chosen to be real. We
neglect any couplings to lighter fermions in our analysis. Aside from
the new couplings, two more free parameters in our model are the dark
matter mass, $m_\chi$, and the mediator mass, $m_{S,P}$. None of them
are explained by the simplified Lagrangian of Eq.~\eqref{eq:lag}, but
could for instance be interpreted in terms of the supersymmetric
realization mentioned above.~\footnote{Unlike for a massive gauge
  boson mediator, the mass of our scalar mediator can be generated
  without additional degrees of freedom and hence without the need to
  extend the simplified model~\cite{bauer}.}
  
%%%%%%%%%%%%%%%%%%%%%%%%%%%%%%%%%%%%%%%%%%%%%%%%%%%%%%%%%%%%%%%%%%%%%%
\subsection{Signals at the LHC}\label{sec:signals}
In the framework defined in Sec.~\ref{sec:model}, we analyze the
different production channels of dark matter at the LHC. Given the
sizable number and precise studies of top-pair events, it seems
promising to search for dark matter produced through an on-shell 
mediator with a large coupling to top quarks in top-pair-associated
production~\cite{cheung_kentarou,chicago,aachen_louvain,haisch_re,ttbar_ex}
\begin{align}
pp \to t\bar{t} \, S/ t\bar{t} \,P 
   \to t\bar{t} \, \chi \bar{\chi} \; .
\end{align}
Using kinematic observables, this production
channel also allows for a determination of some of the mediator
properties~\cite{ttbar_properties,giacomo,pseudo_scalar}. Alternatively,
we can produce the scalar or pseudo-scalar mediator in gluon fusion
through a top loop, with a subsequent decay into a dark matter pair,
\begin{align}
pp \to S X/P X 
   \to \chi \bar{\chi} \, X \; .
\end{align}
In this case, we need to request one or more visible particles $X =
j,Z,\gamma,\dots$ recoiling against the dark matter. While the most
promising of these channels is the generic mono-jet
signal~\cite{monojet,monojet_properties}, the recoiling system can be
as complex as $X=ZH$~\cite{krauss}.  A comparison of top-pair
associated production and mono-jet signals in our scenario 
concludes that the LHC has a better sensitivity to the mono-jet
signal in the current data set~\cite{haisch_re}.

The focus of our work is on dark matter produced in association with a
single top quark from $t$-channel electroweak production,
\begin{align}
pp\to tj\,S/tj\,P 
  \to tj\,\chi\bar\chi \; .
\label{eq:process_dm1}
\end{align}
In the Standard Model, the total cross section of $t$-channel single
top production at the 13 TeV LHC is about a factor of four smaller
than top pair production. If an additional dark matter pair is being
produced, the cross section in the single top channel can be
comparable to the top pair channel. This has been observed in the
contribution of single-top-associated production to the signal region
of top-pair-associated production~\cite{zurich}. A dedicated analysis
of $t$-channel single top production in association with dark matter as a signal
is still missing. While we focus on
$t$-channel single top production \cite{stop_t}, dark matter can in
principle also be produced in association with
$s$-channel~\cite{stop_s} and $W$-associated single top
production~\cite{stop_tw}. Since the $tW$ and $s$-channel production
rates are significantly smaller than $t$-channel production, we expect
the best signal sensitivity from the latter process.

While we were finalizing our study, Ref.~\cite{Pani:2017qyd} appeared,
investigating single top signals of dark fermions that couple to
top quarks through a two-Higgs-doublet model. This model can be
considered an ultra-violet completion of our simplified
model. However, in this case the dominant single top production
channel is $tW$ production. This is due to the presence of a charged scalar mediator. In the limit of heavy charged mediators, the phenomenology of the two-Higgs-doublet model is very similar to our model with a single neutral mediator. \bigskip

Aside from the missing energy signatures, we can also search for
mediators. By construction, the Lagrangian in Eq.~\eqref{eq:lag}
predicts mediator decays into SM particles, resulting in final states
with top-anti-top pairs, as well as two jets or electroweak bosons from
top-loop-induced decays. Since the production of the mediator from
gluon-gluon collisions is loop-suppressed, the sensitivity to
mediators in top-anti-top production is limited, but can be relevant
for light
mediators~\cite{cheung_kentarou,chicago,aachen_louvain,haisch_re,Greiner:2014qna}. Di-jet
resonance searches at the LHC are currently not sensitive to the
region of mediator masses below about $400~\gev$. Di-photon resonance
searches, in turn, can already set limits on the mediator coupling to
top quarks if the decay branching ratio into tops is
sizeable. Searches for signals with four top quarks are not sensitive
yet to scenarios with $\mathcal{O}(1)$ couplings to top quarks, but
will be interesting in the future~\cite{aachen_louvain}. Since
mediator searches probe only the top couplings $g_{S,P}^t$, missing
energy signals are expected to have a better sensitivity to our model
if the ratio $g_{S,P}^\chi/g_{S,P}^t$ is sufficiently large.

%%%%%%%%%%%%%%%%%%%%%%%%%%%%%%%%%%%%%%%%%%%%%%%%%%%%%%%%%%%%%%%%%%%%%%
\subsection{Interpretation as thermal relic}
If we are to interpret our dark matter candidate as a thermal relic,
dark matter annihilation at freeze-out proceeds mostly through the
$s$-channel processes $\chi\bar\chi \to S/P \to gg$ and $\chi\bar\chi
\to S/P \to t\bar t$~\cite{aachen_louvain,dm_eft}. In these
processes, scalar production is suppressed by the relative velocity of
the dark matter particles. In the
pseudo-scalar case, annihilation proceeds through an $S$-wave,
resulting in a smaller relic dark matter abundance in the latter
scenario for fixed model parameters. The $t$-channel process
$\chi\bar\chi \to SS/PP$ is important for $m_{S,P} < m_\chi$. Taking all annihilation processes into account, the
observed relic abundance can be obtained for~\cite{dm_eft}
\begin{align}
10~\gev \lesssim m_{S,P} \lesssim 3 m_\chi \; .
\label{eq:mass_range}
\end{align}
This range assumes a mediator coupling $g_{S,P}^c\times m_c/v$ also to
charm quarks, opening the annihilation channel $\chi\bar\chi \to c
\bar c$. In general, the required couplings are larger for scalar
mediators than for pseudo-scalar mediators.  Below $m_{S,P} = 10~\gev$
constraints from flavor observables are very strong, in particular
when assuming a non-vanishing coupling to bottom
quarks~\cite{Dolan:2014ska}, and cosmological constraints become
relevant. The upper limit lies slightly above the non-relativistic
on-shell condition $m_{S,P} = 2 m_\chi$. At large mediator masses, $m_{S,P} \gg 2m_\chi$, the dark matter annihilation rate becomes strongly suppressed. To avoid an overabundance, we have to invoke another annihilation process in this mass region.\bigskip

At the LHC, for $m_{S,P} < 2 m_\chi$ dark matter production proceeds through an off-shell mediator, resulting in a small production rate. The thermal relic hypothesis is thus difficult to test in missing energy searches in large parts of the mass range identified in Eq.~\eqref{eq:mass_range}. For $m_{S,P} > 2 m_\chi$, on-shell mediator production leads to appreciable rates for the various processes discussed in Sec.~\ref{sec:signals}. This latter case will be in the focus of our analysis.\bigskip

In non-relativistic processes relevant for direct and indirect dark matter detection, scalar and pseudo-scalar mediators
behave very differently. Dark matter-nucleon scattering is induced by
a scalar mediator coupling to gluons via a top-quark
loop~\cite{aachen_louvain}. For a pseudo-scalar mediator, dark
matter-nucleon scattering is
velocity-suppressed~\cite{Agrawal:2010fh}. Constraints on dark matter
from direct detection experiments are thus much weaker for
pseudo-scalar mediators than for scalars. Dark matter annihilation
today results gamma ray spectra from primary or secondary photons. In
our model, gamma ray spectral lines can be created in the
non-relativistic process $\chi\bar\chi \to S/P \to \gamma\gamma$ with
a loop-induced mediator decay. A continuum of gamma rays is produced
from the same annihilation processes governing the thermal relic
density. For scalar mediators, all
annihilation processes are velocity-suppressed in the non-relativistic
limit. Current indirect detection experiments therefore do not
constrain the parameter space of a thermal
relic~\cite{aachen_louvain}. For pseudo-scalar mediators, only the
process $\chi\bar\chi \to PP$ is velocity-suppressed near the
threshold, so that a sizeable flux of photons from the $s$-channel
annihilation processes is expected. In this case, current measurements of gamma rays
from our galactic center and from spheroidal dwarf galaxies are
sensitive to thermal dark matter
candidates~\cite{pseudo_scalar}.

%%%%%%%%%%%%%%%%%%%%%%%%%%%%%%%%%%%%%%%%%%%%%%%%%%%%%%%%%%%%%%%%%%%%%%
\section{Single-top-associated dark matter production}
\label{sec:on_shell}
%
%-------------------------------------------------------------
\begin{figure}[t!]
\begin{center}
\begin{fmfgraph*}(100,60)
\fmfset{arrow_len}{2mm}
\fmfleft{i2,i1}
\fmfright{o5,o4,o3,o2,o1}
\fmf{gluon,width=0.6,lab.side=left,tension=2}{v1,i1}
\fmf{fermion,width=0.6,lab.side=left,tension=1}{o1,v1}
\fmf{fermion,width=0.6,lab.side=right,label=$b$,label.dist=3,tension=2}{v1,v2}
\fmf{fermion,width=0.6,lab.side=left,label=$t^*$,label.dist=3,tension=2}{v2,v4}
\fmf{fermion,width=0.6,lab.side=left,tension=1}{v4,o2}
\fmf{dashes,width=0.6,lab.side=right,label=$S$,label.dist=3,tension=0.6}{v4,v5}
\fmf{fermion,width=0.6,lab.side=left,tension=0.0}{v5,o3}
\fmf{fermion,width=0.6,lab.side=left,tension=1}{o4,v5}
\fmf{fermion,width=0.6,lab.side=left,tension=2}{i2,v3}
\fmf{photon,width=0.6,lab.side=right,label=$W$,label.dist=1,tension=1}{v2,v3}
\fmf{fermion,width=0.6,lab.side=left,tension=1}{v3,o5}
\fmflabel{$g$}{i1}
\fmflabel{$q$}{i2}
\fmflabel{$\bar{b}$}{o1}
\fmflabel{$t$}{o2}
\fmflabel{$\chi$}{o3}
\fmflabel{$\bar{\chi}$}{o4}
\fmflabel{$q'$}{o5}
\end{fmfgraph*}
\hspace*{10mm}
\begin{fmfgraph*}(100,60)
\fmfset{arrow_len}{2mm}
\fmfleft{i2,i1}
\fmfright{o5,o4,o3,o2,o1}
\fmf{gluon,width=0.6,lab.side=left,tension=2}{v1,i1}
\fmf{fermion,width=0.6,lab.side=left,tension=1}{v4,o1}
\fmf{fermion,width=0.6,lab.side=left,tension=1}{v1,v4}
\fmf{fermion,width=0.6,lab.side=left,label=$t$,label.dist=3,tension=2}{v2,v1}
\fmf{fermion,width=0.6,lab.side=left,tension=2}{o4,v2}
\fmf{dashes,width=0.6,lab.side=right,label=$S$,label.dist=3,tension=0.6}{v4,v5}
\fmf{fermion,width=0.6,lab.side=left,tension=0.0}{v5,o2}
\fmf{fermion,width=0.6,lab.side=left,tension=1}{o3,v5}
\fmf{fermion,width=0.6,lab.side=left,tension=2}{i2,v3}
\fmf{photon,width=0.6,lab.side=right,label=$W$,label.dist=1,tension=1}{v2,v3}
\fmf{fermion,width=0.6,lab.side=left,tension=1}{v3,o5}
\fmflabel{$g$}{i1}
\fmflabel{$q$}{i2}
\fmflabel{$t$}{o1}
\fmflabel{$\chi$}{o2}
\fmflabel{$\bar{\chi}$}{o3}
\fmflabel{$\bar{b}$}{o4}
\fmflabel{$q'$}{o5}
\end{fmfgraph*}
\hspace*{20mm}
\begin{fmfgraph*}(100,60)
\fmfset{arrow_len}{2mm}
\fmfleft{i2,i1}
\fmfright{o5,o4,o3,o2}
\fmf{fermion,width=0.6,lab.side=right,label.dist=3,tension=2}{i1,v2}
\fmf{fermion,width=0.6,lab.side=left,label=$t^*$,label.dist=3,tension=2}{v2,v4}
\fmf{fermion,width=0.6,lab.side=left,tension=1}{v4,o2}
\fmf{dashes,width=0.6,lab.side=right,label=$S$,label.dist=3,tension=0.6}{v4,v5}
\fmf{fermion,width=0.6,lab.side=left,tension=0.0}{v5,o3}
\fmf{fermion,width=0.6,lab.side=left,tension=1}{o4,v5}
\fmf{fermion,width=0.6,lab.side=left,tension=2}{i2,v3}
\fmf{photon,width=0.6,lab.side=right,label=$W$,label.dist=1,tension=1}{v2,v3}
\fmf{fermion,width=0.6,lab.side=left,tension=1}{v3,o5}
\fmflabel{$b$}{i1}
\fmflabel{$q$}{i2}
\fmflabel{$t$}{o2}
\fmflabel{$\chi$}{o3}
\fmflabel{$\bar{\chi}$}{o4}
\fmflabel{$q'$}{o5}
\end{fmfgraph*}
\end{center}
\caption{Feynman diagrams describing dark matter production in
  association with a single top quark through the $t$-channel
  process. We show the contributions in the 4-flavor scheme (left two
  diagrams) and in the 5-flavor scheme (right
  diagram), which we use for our simulation.}
\label{fig:feyn}
\end{figure}
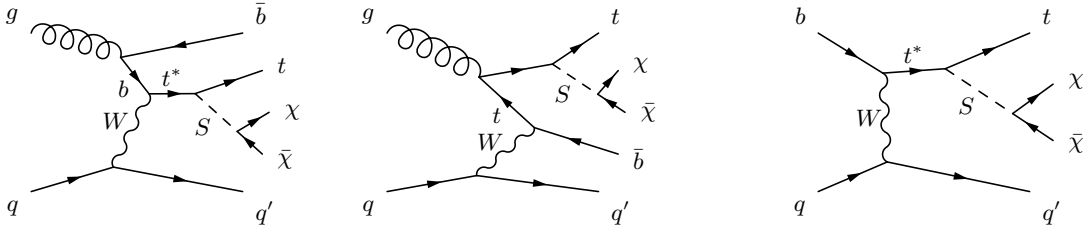
%-----------------------------------
%
In this section, we investigate $t$-channel single top production in
association with a dark matter pair at the LHC.  To maximize the
discovery prospects we focus on a mediator produced on-shell and
decaying into a dark matter pair. We start with the scalar mediator and
discuss the modifications in the pseudo-scalar case in
Sec.~\ref{sec:on_shell_p}. The signal process of single-top-associated
dark matter production can be written as
\begin{align}
pp\to t^* \, j
  \to t \, j  \, S 
  \to t \, j \, (\chi \bar{\chi}) \; .
\label{eq:process_dm3}
\end{align}
Some sample Feynman diagrams are shown in Fig.~\ref{fig:feyn}.  Single
top production in the $t$-channel can be described either in a
4-flavor scheme with incoming gluons splitting into $b\bar{b}$ pairs
or in a 5-flavor scheme, where the bottom-quark is considered as a
parton inside the proton. The difference between the 4-flavor and
5-flavor approaches is the treatment of collinear logarithms in the
perturbative QCD series and can be moderated by including higher-order
QCD corrections~\cite{bottom_density}. For our simulation, we use the
5-flavor scheme with its resummation-improved total cross section.\bigskip

%-----------------------------------
\begin{figure}[t]
\includegraphics[width=0.48\hsize]{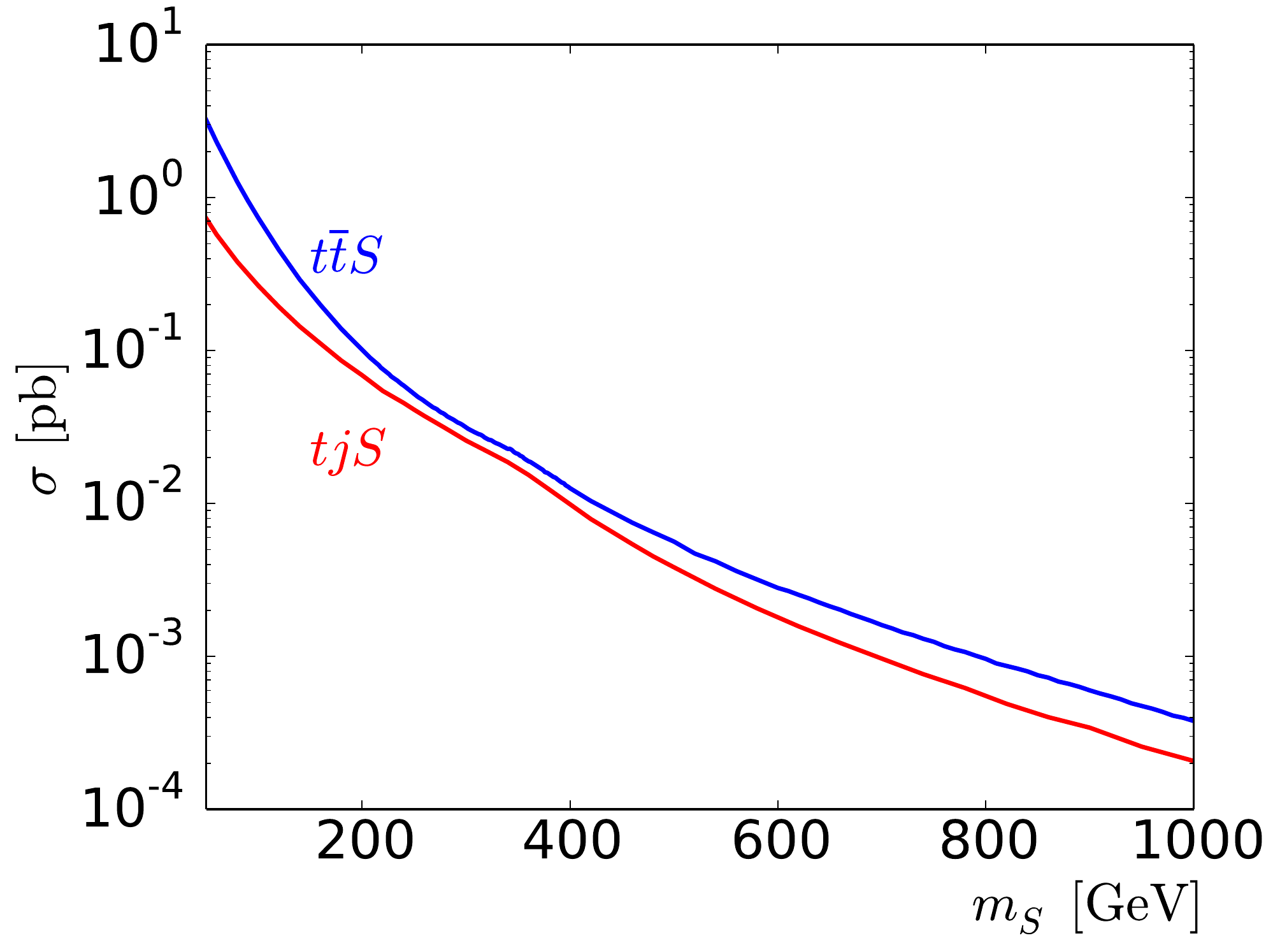}
\hspace*{0.05\hsize}
\includegraphics[width=0.45\hsize]{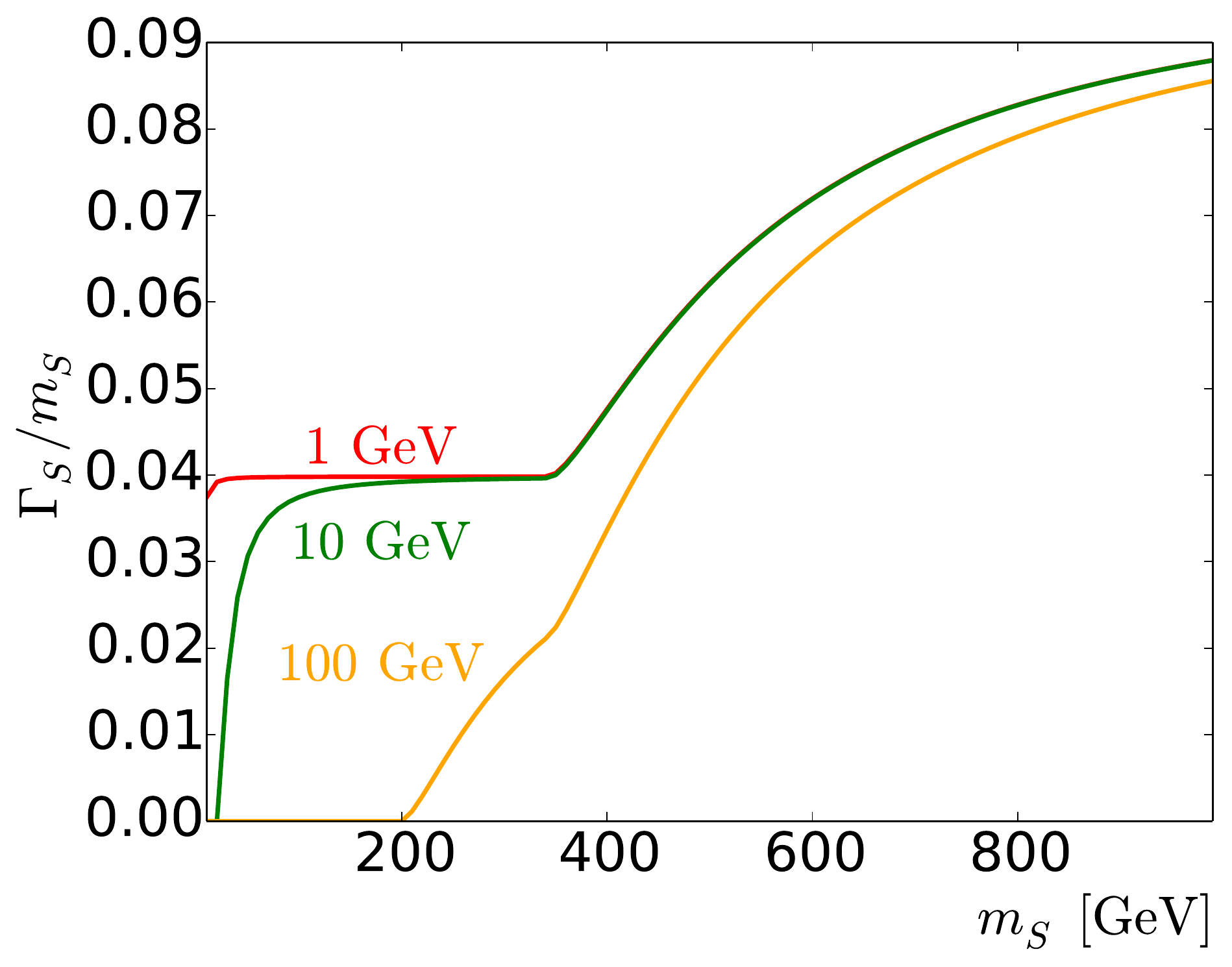}
\caption{Left: production rate for single-top-associated (red) and
  top-pair-associated (blue) dark matter production with an on-shell
  scalar mediator. Right: width-to-mass ratio for the scalar
  mediator. We assume $g_S^t=g_S^\chi=1$ and a default mass value of $m_\chi =
  1\,\text{GeV}$.}
\label{fig:rate_s}
\end{figure}
%-----------------------------------

At the LHC, a heavy mediator with $m_{S,P} > 2m_\chi$ is produced
on-shell, such that dark matter production factorizes into resonant
mediator production and subsequent decay to a dark matter pair. According to the Lagrangian in Eq.~\eqref{eq:lag}, the relevant model
parameters for the mediator production are the mediator mass, $m_S$,
and the top coupling, $g_S^t$. In addition, the total rate of dark
matter production depends on the branching ratio of the mediator into
the dark matter pair. The decay width of the mediator width is given
to a good approximation by the sum of the partial decay rates into
$\chi\bar\chi$ and $t\bar t$ final states,
\begin{align}
\frac{\Gamma_S}{m_S}
&= \frac{1}{8\pi}\Bigg[(g_S^{\chi})^2 \left(1-\frac{4m_\chi^2}{m_S^2}\right)^{3/2} 
   + 3 (g_S^t)^2 \frac{m_t^2}{v^2} \left(1-\frac{4m_t^2}{m_S^2}\right)^{3/2} \Theta(m_S - 2m_t) \Bigg].
\label{eq:scalar_w}
\end{align}
Loop-induced mediator decays into pairs of gluons or photons are
numerically subleading for on-shell mediators and thus neglected in
our analysis. In the right panel of Fig.~\ref{fig:rate_s}, we show the
total decay rate $\Gamma_S/m_S$ of the mediator for fixed dark matter
masses $m_\chi=1,10,100~\gev$. For $m_S < \text{min} (2m_\chi, 2m_t)$,
the mediator can only decay via loop-suppressed processes, resulting
in a very narrow resonance.  For $2m_\chi < m_S < 2m_t$, the width is
dominated by the decay into $\chi \bar\chi$ alone. Because the
branching ratio $\mathcal{B}(S\to \chi\bar\chi)$ is close to one, dark
matter production via an on-shell mediator is essentially independent
of the dark matter mass and coupling. Above the top threshold the
decay into $\chi\bar\chi$ competes with $t\bar t$, so that dark matter pair
production will depend on $m_\chi$ and $g_S^\chi$ through the
mediator width. For mediator masses up to the TeV scale the width
remains narrow, $\Gamma_S/m_S \lesssim 9\,\%$, and the resonance is
described by a Breit-Wigner propagator.

As the default model setup for our analysis we choose a benchmark scenario for which we expect a high sensitivity in single-top-associated dark
matter production,
\begin{align}
g_{S,P}^t = g_{S,P}^\chi = 1,\qqquad m_\chi = 1~\gev,\qqquad m_{S,P} = 300~\gev \; ,
\label{eq:benchmark}
\end{align}
We also use it to be able to compare with
Refs.~\cite{ttbar_properties,giacomo,pseudo_scalar}. As long as the mediator is produced on-shell, results for a different dark matter mass can be
deduced rather easily, because it only enters the signal indirectly through the mediator width. For mediator masses $m_{S,P} < 2m_\chi$, the mediator is off-shell, so that we need to consider the full process with two dark matter particles in the final state. In this case, the production rate at the LHC is much smaller, due to the lack of the resonance enhancement. For mediators around the electroweak scale, $tj\chi\bar{\chi}$ off-shell rates range about two orders of magnitude below the rates with a resonant mediator. Kinematic distributions, in turn, look very similar with on-shell and off-shell mediators. Since single-top-associated production is not sensitive to the off-shell scenario with perturbative couplings, we will focus on on-shell mediator production in what follows.

In the left panel of Fig.~\ref{fig:rate_s}, we show the dark matter
production rates for our on-shell benchmark scenario with a scalar mediator from Eq.~\eqref{eq:benchmark}. We compare associated production with a single top quark, $pp\to t j S \to t j (\chi\bar \chi)$, and a top-anti-top pair,
$pp\to t \bar t\,S \to t\bar t\,(\chi\bar \chi)$. The single top rate becomes comparable to top-anti-top for scalar
mediator masses $200~\gev \lesssim m_S \lesssim 500~\gev$. Compared with the SM
predictions, the additional radiation of a heavy mediator favors
incoming quarks over incoming gluons and thus single top over top pair
production. In Tab.~\ref{tab:signalxs}, we show the cross
section for top pair and single top production in the Standard Model
and in association with dark matter for our benchmark scenario from Eq.~\eqref{eq:benchmark}. The large difference between single top and top pair production observed in the Standard Model is clearly lifted, once a heavy mediator is radiated from the top quark. Further features of the scalar production rate will be discussed in comparison
with the pseudo-scalar in Sec.~\ref{sec:on_shell_p}.

%%%%%%%%%%%%%%%%%%%%%%%
\begin{table}
\begin{tabular}{l | r || l | r}
\ Standard Model & $\ \sigma_\text{tot}$ [pb] $\ $ & $\ $ dark matter signal $\ $ & $\ \sigma_\text{tot}$ [fb] \\
\hline
$\ pp\to t\bar t$ & $\ $ $832$~\cite{Czakon:2013goa} $\ $ & $\ $ $pp\to t\bar t \, \chi\bar \chi$ & $30$ $\ $ \\
\hline
$\ pp\to tj$ ($t$-channel) & $\ $ $217$~\cite{Kant:2014oha} $\ $ & $\ $ $pp\to tj \,  \chi\bar \chi$  & $26$ $\ $ \\
$\ pp\to tW$ & $\ $ $\ 72$~\cite{Kidonakis:2015nna} $\ $ & $\ $ $pp\to tW \,\chi\bar \chi$  & $9$ $\ $ \\
$\ pp\to tj_b$ ($s$-channel) $\ $ & $\ $ $10$~\cite{Kant:2014oha} $\ $ & $\ $ $pp\to tj_b \, \chi\bar \chi$  & $0.01$ $\ $ \\
\end{tabular}
\caption{Total cross sections for top pair and single top production in the Standard Model and in association with a dark matter pair with a scalar mediator for $g_S^t = g_S^\chi = 1$, $m_\chi = 1~\gev$, $m_S = 300~\gev$ at the 13-TeV LHC. For all single top processes, we quote the sum of top and anti-top production.}
 \label{tab:signalxs}
\end{table}
%%%%%%%%%%%%%%%%%%%%%%%

%%%%%%%%%%%%%%%%%%%%%%%%%%%%%%%%%%%%%%%%%%%%%%%%%%%%%%%%%%%%%%%%%%%%%%
\subsection{Signal extraction for a scalar mediator}
\label{sec:on_shell_s}
For our signal versus background analysis we focus on the leptonic decay
of the top quark to ensure that the signal passes standard
triggers. An important characteristic of the single top channel is a
 forward light-flavor jet.  The final state consists of exactly one
lepton $\ell = e,\mu$, exactly one $b$-tagged jet, at least one
light-flavor jet, and a significant amount of missing transverse
energy,
\begin{align}
pp\to tj\chi\bar\chi \to (\ell b) j + \met,
\end{align}
In all our results, we consider the sum of top and anti-top quarks in
single top production.  We simulate all signal and background events
for the 13~TeV LHC with \textsc{Sherpa}~\cite{sherpa} and its \textsc{Ufo}
interface to FeynRules~\cite{feynrules,ufo,sherpaufo}.  Our
simulations are performed at the leading order in QCD and include the
parton shower. For the reconstruction of anti-$k_T$ jets with $R=0.4$
we rely on \textsc{FastJet}~\cite{fastjet}.  We use the set of parton
distributions \textsc{Nnpdf3.1lo} with
$\alpha_s(m_Z)=0.118$~\cite{Ball:2017nwa}. The top-quark mass is set
to $m_t = 172.0~\gev$. To obtain realistic identification
efficiencies, we include the fast detector simulation
\textsc{Delphes}~\cite{deFavereau:2013fsa} with the characteristics of
the ATLAS detector.

%-----------------------------------
\begin{figure}[t]
\includegraphics[width=0.45\hsize]{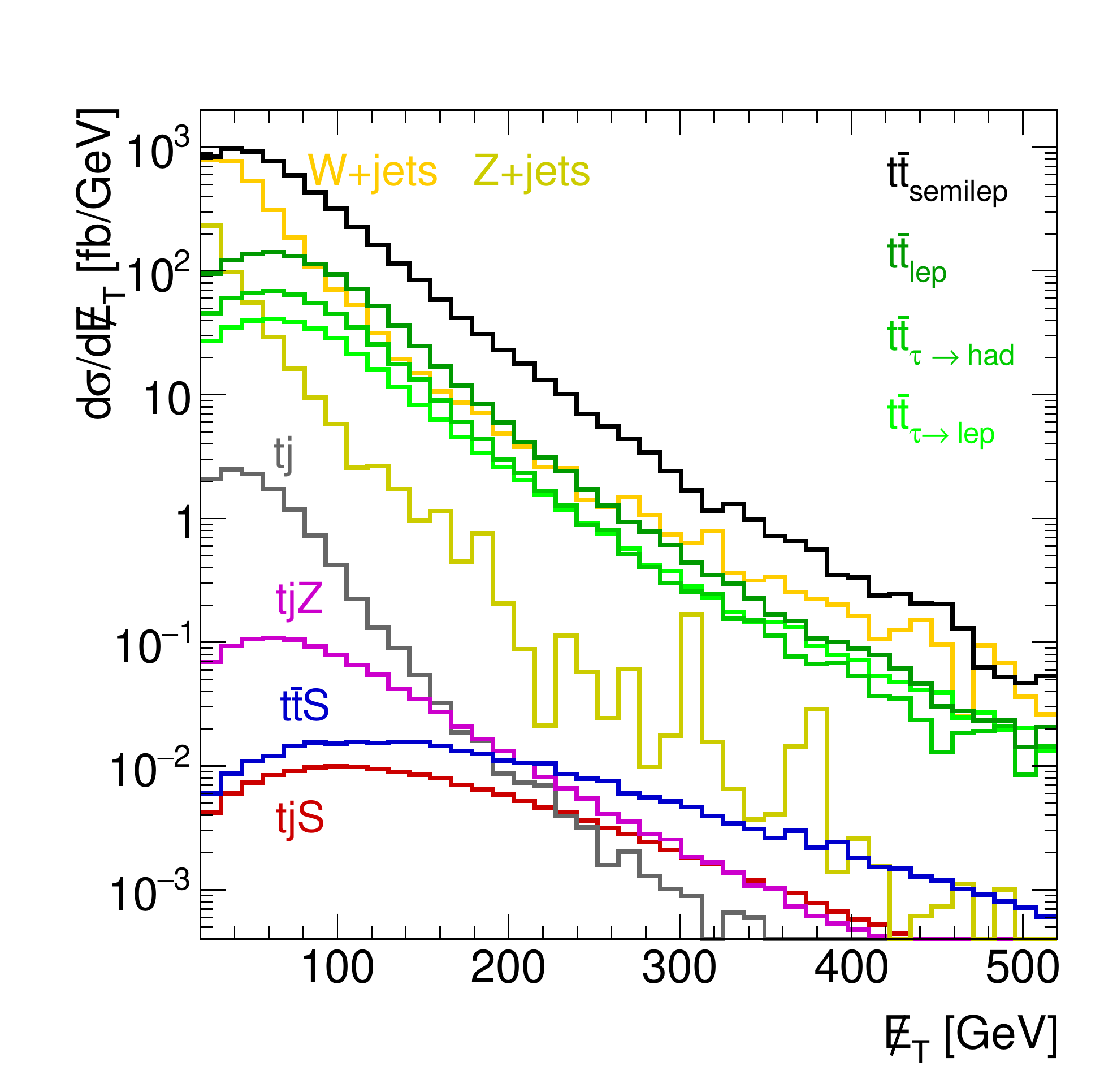}
\hspace*{0.05\hsize}
\includegraphics[width=0.45\hsize]{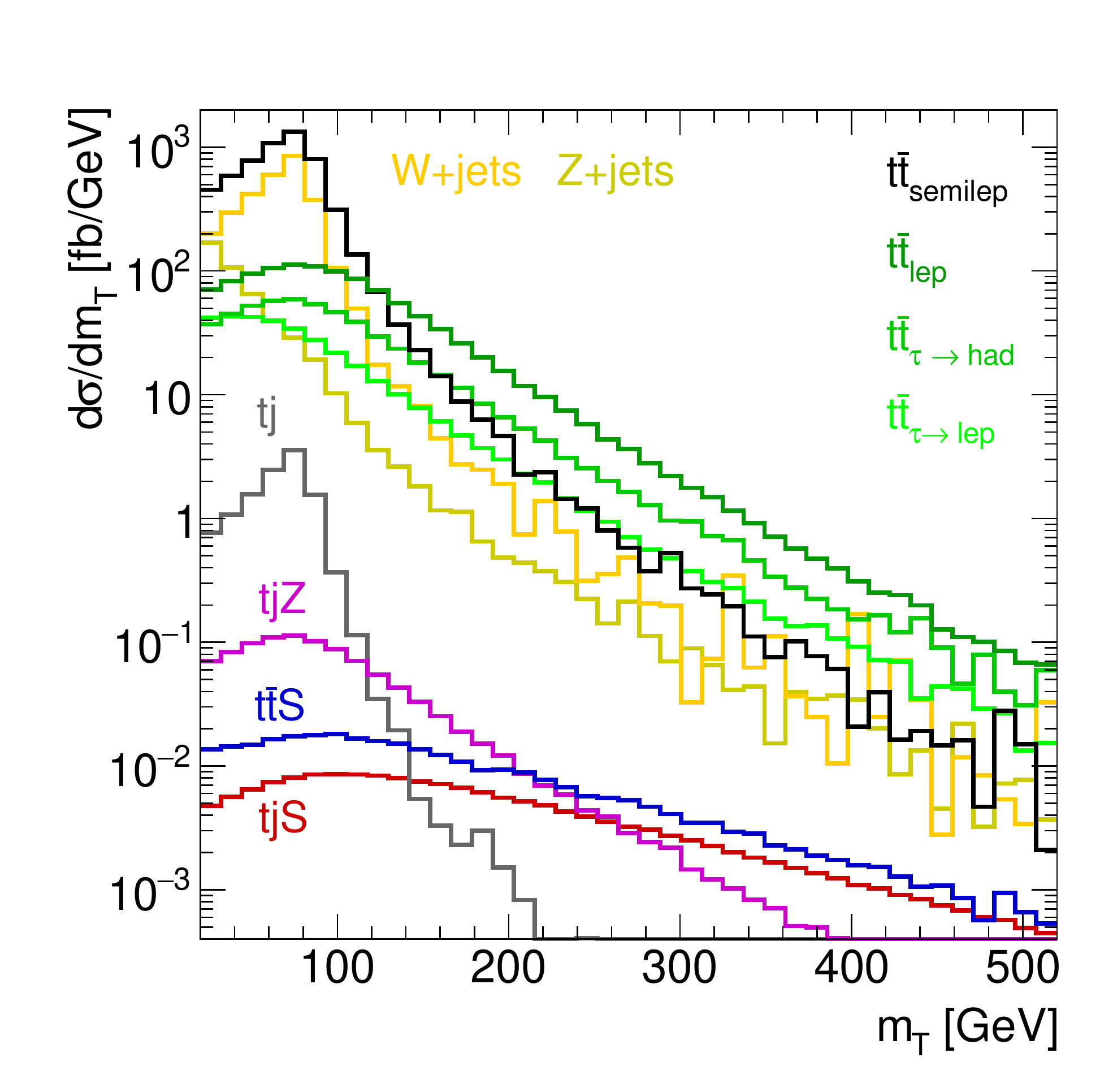} 
\caption{Kinematic distributions for the signal process $pp \to tjS \to tj\chi\bar{\chi}$ with a scalar mediator and the relevant backgrounds: the missing transverse energy $\met$ (left) and the transverse mass $m_T$ of the lepton and missing
  momenta (right). Distributions are shown for the benchmark parameters in Eq.~\eqref{eq:benchmark} and after applying the cuts from Eq.~\eqref{eq:acceptance1}.}
\label{fig:distri_s1}
\end{figure}
%-----------------------------------

For the signal we employ the 5-flavor scheme, setting the factorization scale to $\mu_F = 200\,\text{GeV}$. We have checked that the kinematics of all final-state
particles except for the spectator $b$-jet are identical to the
4-flavor scheme.  Since our dark sector is color-blind, higher-order
QCD corrections to our signal process should be similar to $t$-channel
single top production in the Standard Model, where they do not exceed
a few percent~\cite{Brucherseifer:2014ama}.\bigskip

We begin our analysis with a set of loose acceptance cuts 
\begin{alignat}{9}
p_{T,\ell}   &> 20~\gev, &\qqqquad 
|\eta_\ell |, |\eta_b| &< 2.5, \notag \\
p_{T,b},\,p_{T,j} &> 20~\gev, &\qqqquad 
|\eta_j| &< 4.5,  
\label{eq:acceptance1}
\end{alignat}
If there is more than one non-$b$-tagged jet, we use the hardest jet
in $p_T$.  After applying the above acceptance cuts, the dominant backgrounds are
\begin{itemize}
\setlength{\itemsep}{0pt}
  \item[a) ] $t\bar{t}$ production, with one leptonically and one
    hadronically decaying top. Alternatively, both tops can decay
    leptonically, but in this case one lepton is missed by the lepton
    veto. The missing energy then comes from a combination of two
    neutrinos. Events with one or both tops decaying into tau leptons
    with subsequent hadronic decays result in final states with larger
    missing energy, due to the presence of an additional neutrino.
    We simulate the top-pair background at LO QCD with one merged hard
    jet~\cite{ckkw} and normalize the rate to the NNLO QCD 
    prediction from Tab.~\ref{tab:signalxs}.
\item[b) ] $tW$ and $tZ$ production with a leptonically decaying top
  quark and actual or fake missing energy from the gauge boson
  decay. For heavy mediators or dark matter pairs, the spectrum of
  missing transverse energy is softer than for the single top signal
  and can be used to reject these backgrounds.
\item[c) ] $t$-channel single top production, which can be efficiently rejected
  due to its softer $\met$ spectrum and a steep drop-off at high
  transverse masses (see below).
\item[d) ] $W$+jets production with a true or fake $b$-jet, which also
  features a strong drop-off in the transverse mass spectrum.
  $Z$+jets production with a missing lepton is small in comparison.
\item[e) ] $t\bar{t} \chi \bar{\chi}$ production, which can be
  efficiently rejected by requesting the hardest light-flavor jet to
  be emitted in the forward region and by exploiting its kinematic
  correlation with the $b$-jet (see below). An efficient suppression
  of this background is necessary to eventually combine both
  channels~\cite{zurich}.
\end{itemize}

%-----------------------------------
\begin{figure}[t]
\includegraphics[width=0.45\hsize]{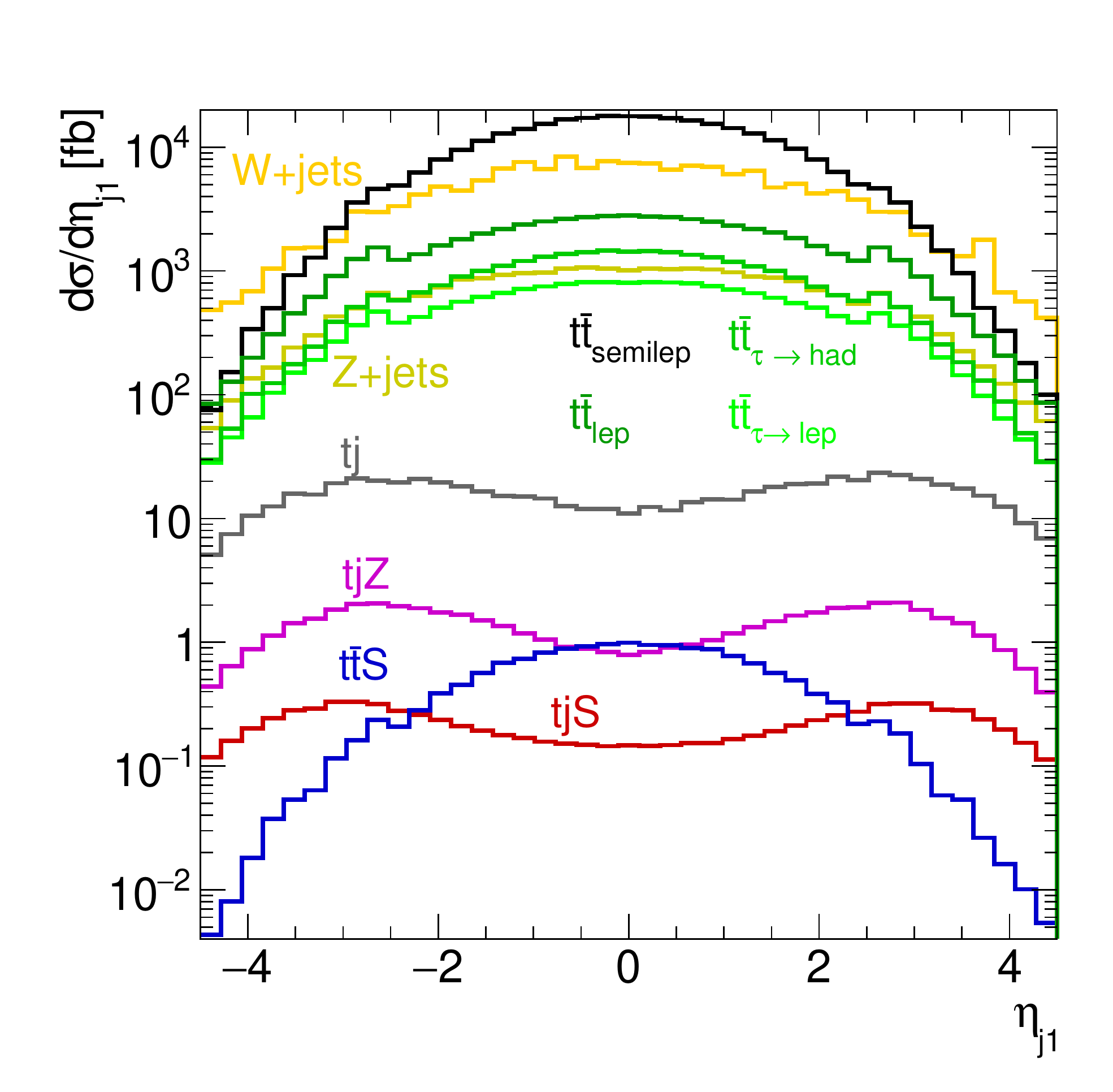}
\hspace*{0.05\hsize}
\includegraphics[width=0.45\hsize]{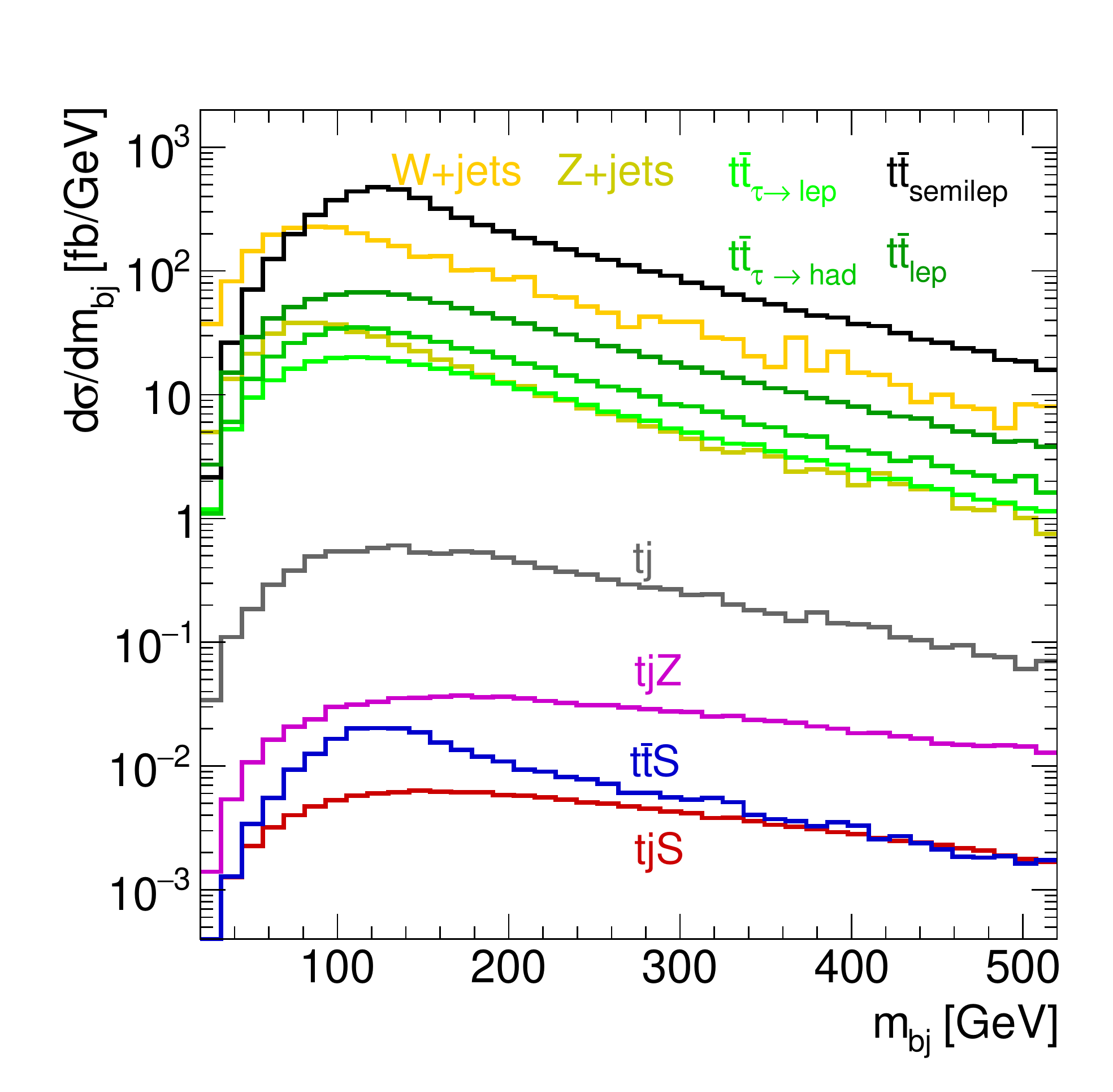}
\caption{Signal and background distributions of the hardest
  light-flavor jet rapidity (left) and the invariant mass $m_{bj}$ (right) in the benchmark scenario from Eq.~\eqref{eq:benchmark} with a scalar mediator and after applying the acceptance cuts from Eq.~\eqref{eq:acceptance1}.}
\label{fig:distri_s3}
\end{figure}
%-----------------------------------

Kinematic observables help discriminating between signal and
background. We discuss their impact based on our scalar benchmark
model from Eq.~(\ref{eq:benchmark}). In the left panel of
Fig.~\ref{fig:distri_s1}, we show distributions of the missing
transverse energy for the signal $tjS(\to \chi\bar{\chi})$ and the
backgrounds mentioned above. The signal leads to a significantly
harder missing energy spectrum, especially if the produced mediator is
heavy. The dominant backgrounds are $t\bar t$ and $W$+jets production,
while all other backgrounds are strongly suppressed at high
$\met$. Notice that the $t\bar{t}S(\to \chi\bar{\chi})$ signal
produces a hard spectrum of missing energy as well. As we will show
below, additional kinematic observables can efficiently reduce its
contribution to the signal region of $tjS(\to
\chi\bar{\chi})$.\bigskip

In the $tjS$ signal, the missing energy comes from a combination of
the neutrino and mediator momenta, while in the backgrounds it is
typically generated by neutrinos from $W$ decays. We exploit this
feature through the transverse mass $m_T$ of the
lepton and the sum of missing particles' momenta defined by
\begin{align}
m_T^2 = 2p_{T,\ell} \met (1-\cos \phi_{\ell,\met}).
\end{align}
Here $\phi_{\ell,\met}$ is the azimuthal angular separation of the lepton and missing momenta. In the right panel of Fig.~\ref{fig:distri_s1},
we confirm that for backgrounds with one neutrino from a $W$ decay
there is a cliff around $m_T \approx M_W$. Off-shell effects, width
effects, combinatorics, and detector effects lead to a low number of
remnant events above this threshold. This allows us to drastically
reduce the single-lepton $t\bar t$ background, the single top
background, and $W$+jets. Therefore, we extend the pre-selection cuts shown in Eq.~\eqref{eq:acceptance1} by
\begin{align}
\met > 200~\gev , \qquad \text{and} \qquad m_T > 85~\gev \; .
\label{eq:acceptance2}
\end{align}

It is well known that $t$-channel
single top production leads to a hard jet in the forward
region~\cite{singletop_jets}. In the left panel of
Fig.~\ref{fig:distri_s3}, we show the rapidity distribution of the
hardest light-flavor jet for the signal and the backgrounds. In particular for the
$t\bar t$ backgrounds, the jet indeed tends to be much more central
than for the signal. As a third pre-selection cut we thus require
\begin{align}
|\eta_{j_1}| > 2 \; .
\label{eq:acceptance3}
\end{align}
The jet rapidity is also a useful discriminator between the $tjS$ and $t\bar{t}S$
signals. Besides the rapidity, we furthermore exploit
the kinematic correlations of the light-flavor jet and the $b$-jet. In
the right panel of Fig.~\ref{fig:distri_s3}, we see that the invariant
mass $m_{bj}$ is large for the signal. This observable becomes more
distinctive when the mediator $S$ is heavy.\bigskip

The combination of Eqs.~\eqref{eq:acceptance1}, \eqref{eq:acceptance2}, and \eqref{eq:acceptance3} defines the pre-selection before we enter a dedicated analysis.
 At this stage, the signal rate for our model
benchmark from Eq.~\eqref{eq:benchmark} is
\begin{align}
\sigma_{t_\ell j \, \chi\bar\chi} = 0.2~\text{fb},
\label{eq:signal_rate}
\end{align}
including the leptonic branching ratio of the top quark.  The relevant
backgrounds after pre-selection are
\begin{alignat}{9}
\sigma_{t_\ell \bar t_\ell} = 16.8~\text{fb}, \qquad 
\sigma_{t_\ell \bar t_{\tau_h}} = 6.2~\text{fb}, \qquad
\sigma_{t_\ell \bar t_{\tau_\ell}} = 6.1~\text{fb}, \qquad
\sigma_{t_\ell \bar t_h} = 5.1~\text{fb} \; .
\label{eq:background_rate}
\end{alignat}
Since we consider the sum of top and anti-top quarks in our single top
signal, $t\bar t$ rates with distinguishable top and anti-top decays
contribute twice to the background.  The two cuts in
Eq.~\eqref{eq:acceptance2}, in particular the cut on the transverse
mass, offer excellent opportunities to define background control
regions. The number of $t\bar{t}$ background events in the signal
region defined by our pre-selection is about $10^{-5}$ of the
full top pair production sample, allowing for a solid statistical
coverage even of suppressed phase space regions, for instance with sizable
$\met$. This implies that the background estimate in the signal region
will be dominated by systematic uncertainties from the background
extrapolation for well-understood processes like top pair production
and $W$+jets or $Z$+jets production. More challenging backgrounds, like
$t j Z$ production, are clearly sub-leading, as can be
seen in Fig.~\ref{fig:distri_s1}.\bigskip

In addition to the rather general observables discussed above, we can target
specific backgrounds with high-level kinematic observables. As a
starting point, whenever the neutrino originates from a top decay, we
can complement the assumed lepton-neutrino transverse mass $m_T$ with
the transverse mass of the bottom-lepton-$\met$ system and require the
latter to be larger than $m_t$. This targets specifically the $t\bar
t$ background with one leptonically and one hadronically decaying
top.

For backgrounds with two or more neutrinos the distributions in
Fig.~\ref{fig:distri_s1} look very similar to the signal.  The same is
true for $t\bar t \chi\bar{\chi}$ production.  To reject $t\bar t$
production with two leptonically decaying tops and one lepton missed,
we use a dedicated variable that fully exploits the kinematic topology
of this background~\cite{mt2},
\begin{alignat}{9}
M_{T2}^W = \min_{\vec p_1 + \vec p_2 = \slashchar{\vec{p}}_T} \tilde{m}_t
\qqquad \text{with} \quad 
p_1^2&=0 &\qquad &\text{(assumed neutrino)} \notag \\[-3mm]
(p_1 + p_\ell)^2 &= m_W^2 &\qquad &\text{(assumed $W$ with detected lepton)} \notag \\
(p_1 + p_\ell + p_{b,1})^2 &= \tilde{m}_t^2 &\qquad &\text{(assumed $t$ with detected lepton)} \notag \\
p_2^2&= m_W^2 &\qquad &\text{(assumed $W$ with missed lepton)} \notag \\
(p_2 + p_{b,2})^2 &= \tilde{m}_t^2 &\qquad &\text{(assumed $t$ with missed lepton).}
\end{alignat}
Here $p_1$ is assumed to be the momentum of the neutrino paired with
the detected lepton, whereas $p_2$ is the sum of
momenta from the other neutrino and missed lepton. Since our signal
features exactly one $b$-tagged jet, for the other $b$-momentum we use
the momentum of a light-quark jet (stemming from a mis-identified
$b$-jet).  If more than one light-quark jet is observed, we take the
value of the hardest or second-hardest jet in $p_T$, which gives the
smaller value of $M_{T2}^W$. For large mediator masses, $M_{T2}^W$ is an efficient discriminator
between signal and purely leptonic top-pair background. At lower
mediator masses, the signal and top-pair topologies look more similar
and the discriminating power of the $M_{T2}^W$ variable is
reduced.

%%%%%%%%%%%%%%%%%%%%%%%%%%%%%%%%%%%%%%%%%%%%%%%%%%%%%%%%%%%%%%%%%%%%%%
\subsection{Multi-variate analysis}
\label{sec:multivariate}
Since the single top signal differs from the background channels in
many kinematic observables, and because the signal rate given in
Eq.~\eqref{eq:signal_rate} is small, we employ a multi-variate method
to separate signal and background regions in phase space. We use
boosted decision trees (BDT) in \textsc{Tmva}~\cite{tmva} after the
pre-selection cuts of Eq.~\eqref{eq:acceptance1}, 
Eq.~\eqref{eq:acceptance2} and Eq.~\eqref{eq:acceptance3}. The input variables describing the lepton,
$b$-jet and light-flavor jet in final state, as well as the missing
transverse momentum vector, are
\begin{align}
\left\{ p_{T,\ell}, \eta_{\ell}, \; p_{T,b}, \eta_{b}, \; p_{T,j_1}, \eta_{j_1}, \; \met, \;
        \phi_{\ell,b}, \phi_{\ell,j_1}, \phi_{j_1,b}, \phi_{\ell,\met},  \phi_{j_1,\met}, \phi_{b,\met}, \; 
        m_T, M_{T2}^W, m_{bj_1}, \; N_\text{jets}  \right\} \; .
\label{eq:multivariate}
\end{align}
Here $\phi_{m,n}$ denotes the azimuthal angle between objects $m$ and
$n$, and $N_{\text{jets}}$ is the number of detected light-quark
jets. We expect that at the LHC the uncertainty of the analysis will
be statistics dominated. Due to the large number of background events,
powerful control regions are important to obtain a high signal
sensitivity. Based on our discussion above, we assume a remaining systematic
uncertainty of 3\% or at most 10\% on the combined backgrounds in our analysis.  This
relative systematic uncertainty from the background extrapolation is
much smaller than the background uncertainty quoted for the
$t\bar{t}\chi\bar{\chi}$ analysis in Ref.~\cite{giacomo}. In the latter analysis, the leading background is $t\bar t Z$ production, while we quote
our uncertainty relative to the leading $t\bar{t}$ background. This
corresponds to the key difference between our analysis and
Ref.~\cite{giacomo}: We do not attempt to entirely remove the
background through cuts to define appropriate signal regions.

The LHC reach for our model depends on the dark matter couplings
$g_S^t$ and $g_S^\chi$ defined in Eq.~\eqref{eq:lag} and the mediator
mass $m_S$. Assuming $m_\chi \ll m_S$, the signal rate below and above
the threshold for mediator decays to top pairs roughly scales like
\begin{alignat}{9}
\sigma_{tj\chi\bar{\chi}} &\propto |g_S^t|^2 &\qqqquad
&m_\chi \ll m_S < 2 m_t \; , \notag \\
\sigma_{tj\chi\bar{\chi}} &\propto |g_S^\chi|^2\left( 3\frac{m_t^2}{v^2}\Big(1-\frac{4m_t^2}{m_S^2}\Big)^{3/2} + \frac{|g_S^\chi|^2}{|g_S^t|^2} \right)^{-1}  & \qqqquad 
&m_\chi \ll 2 m_t < m_S \; .
\label{eq:scale-rates}
\end{alignat}
For heavier mediators we observe an additional suppression through the
total mediator width.

%-----------------------------------
\begin{figure}[t]
\includegraphics[width=0.46\hsize]{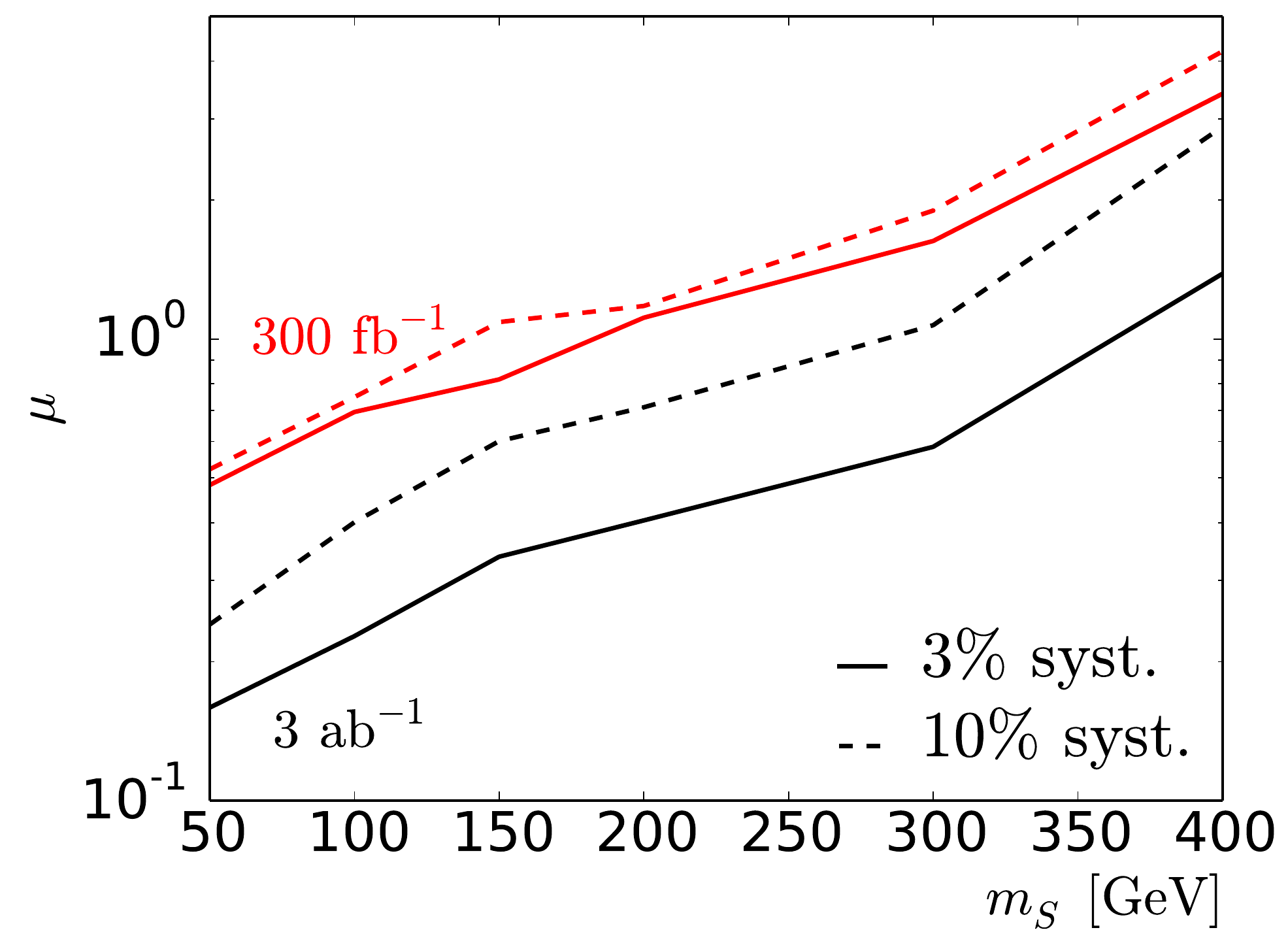} 
\hspace*{0.05\hsize} 
\includegraphics[width=0.46\hsize]{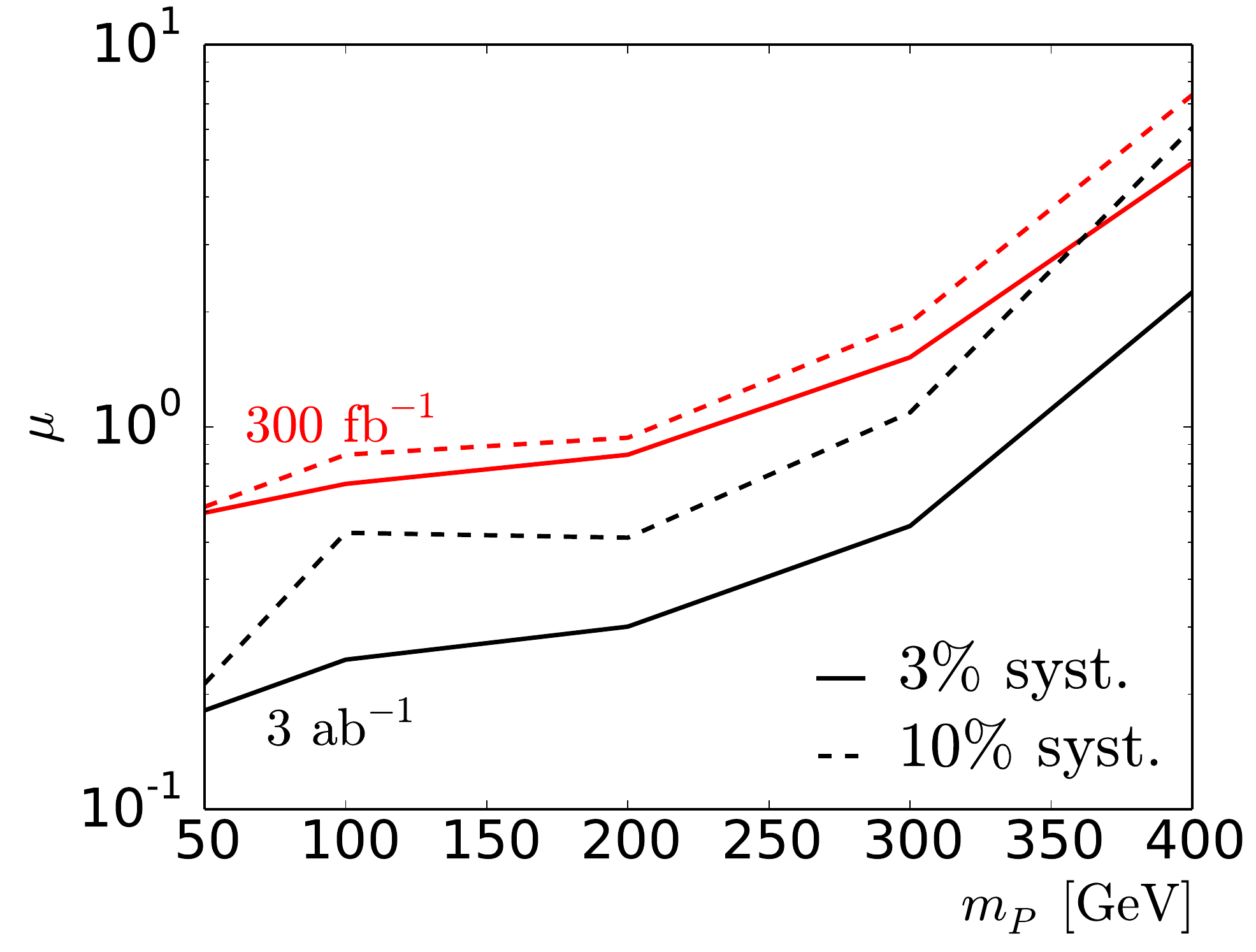}
\caption{Sensitivity to single-top-associated dark matter production with a scalar (left) and a pseudo-scalar (right)
  mediator at the 13 TeV LHC with $300\,\text{fb}^{-1}$ (red) and
  $3\,\text{ab}^{-1}$ (black), assuming a systematic background uncertainty of $3\%$ (plain) and $10\%$ (dashed). Shown is the signal strength $\mu$ that can be excluded at $95\%$ CL, as a function of the mediator mass.}
\label{fig:significance_s}
\end{figure}
%-----------------------------------
 
The sensitivity to our signal is parameterized in terms of the signal strength $\mu$, defined as the ratio of observed
events in our pre-selection region over the expected event rate for scalar couplings
$g_S^t=g_S^\chi=1$ and dark matter mass $m_\chi=1\,\text{GeV}$ (and likewise for a pseudo-scalar mediator). In Fig.~\ref{fig:significance_s}, we show the expected signal strength that can be excluded at the $95\%$ confidence level (CL) with $300\,\text{fb}^{-1}$ (red) and
$3\,\text{ab}^{-1}$ (black) of data, assuming a systematic background uncertainty of $3\%$ (plain) and $10\%$ (dashed), respectively. 

To obtain these results, we have used the $\text{CL}_S$ method~\cite{cls} implemented in
\textsc{CheckMate}~\cite{checkmate}, interpreting the obtained values
of $\text{CL}_S < 0.05$ as excluded at the $95\%$ CL. In this approach
we apply a variable cut on the BDT output parameter to define the
signal region and an associated set of signal and background
rates. This is the same cut as one would apply to construct an ROC
curve describing the signal versus background efficiencies of the
multi-variate analysis. For each value of this cut on the BDT output
we compute the reach in terms of the $\text{CL}_S$ value and then
quote the best possible limit. On the one hand, this method does not provide the
best limit one can reach, because it does not use the actual shape of
the BDT output curve and is therefore hardly sensitive to small but
highly distinctive regions of phase space. On the other hand, it
avoids giving too much weight to the systematics-limited tails of
kinematic distributions. Instead, the integrated rate above a given
cut on the BDT output parameter will include a large number of events
from the statistics-limited $t\bar{t}$ background in all kinematic
distributions. Larger uncertainties on sub-leading background
processes will not have any sizeable impact on our numerical
results.

In Fig.~\ref{fig:significance_s}, we see how the expected LHC sensitivity in
the region $m_\chi \ll m_S < 2 m_t$ drops rapidly for larger mediator masses. This is due to an overall reduced production rate, as well as changing signal kinematics. At light mediator masses, the
most powerful signal features are the rapidity of the forward jet, the
missing transverse momentum, and the number of jets in the event.  A
heavier mediator is produced closer to threshold and generates less
missing transverse energy. Similarly, a larger mediator mass increases the collinear
logarithm dominating the kinematics of the forward light-flavor jet
shown in Fig.~\ref{fig:feyn}, pushing it to larger rapidities. This effect changes the impact of kinematic variables, such as the azimuthal angle between the lepton and the missing transverse momentum, which gains relevance for larger mediator masses. The
net result of these kinematic features is that the sensitivity is
significantly reduced.

Using Eq.~\eqref{eq:scale-rates}, the signal strength directly translates into a bound on the mediator coupling to top
quarks. For a scalar mass $m_S = 50\,\text{GeV}$, couplings $g_S^t \gtrsim 0.7 (0.4)$ can
be excluded at $95\%$ CL by the LHC with $300\,\text{fb}^{-1}$
($3\,\text{ab}^{-1}$). The sensitivity is comparable with that expected from top-pair-associated production~\cite{giacomo}. A signal strength of $\mu=1$ can ultimately be probed at the LHC with $3\,\text{ab}^{-1}$ for masses $m_S <
360\,\text{GeV}$. For comparison, in top-pair-associated production the reach
extends to $m_S < 400\,\text{GeV}$.

%%%%%%%%%%%%%%%%%%%%%%%%%%%%%%%%%%%%%%%%%%%%%%%%%%%%%%%%%%%%%%%%%%%%%%
\subsection{Pseudo-scalar versus scalar mediators}
\label{sec:on_shell_p}
%
%-----------------------------------
\begin{figure}[t]
\includegraphics[width=0.45\hsize]{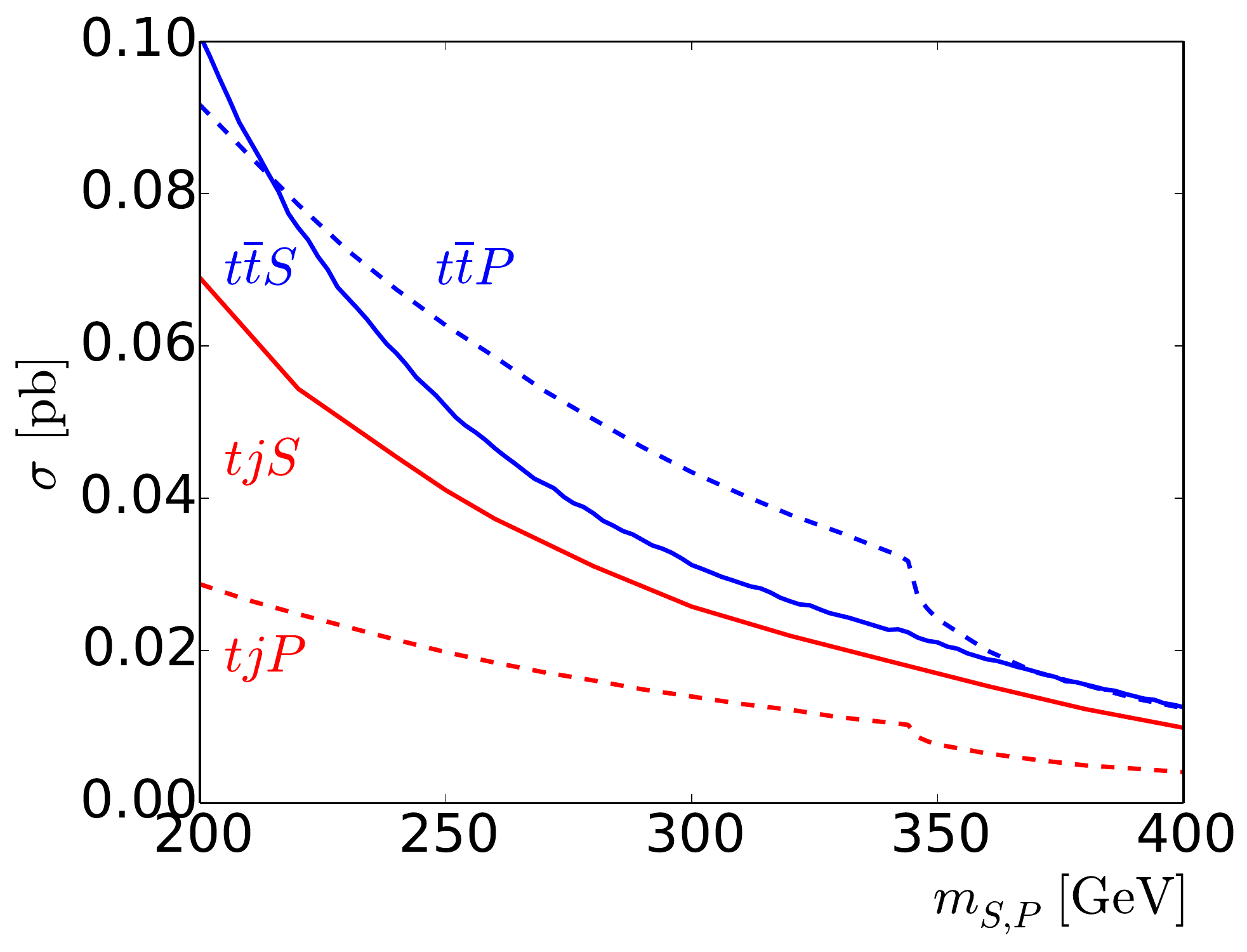}
\hspace*{0.05\hsize}
\includegraphics[width=0.45\hsize]{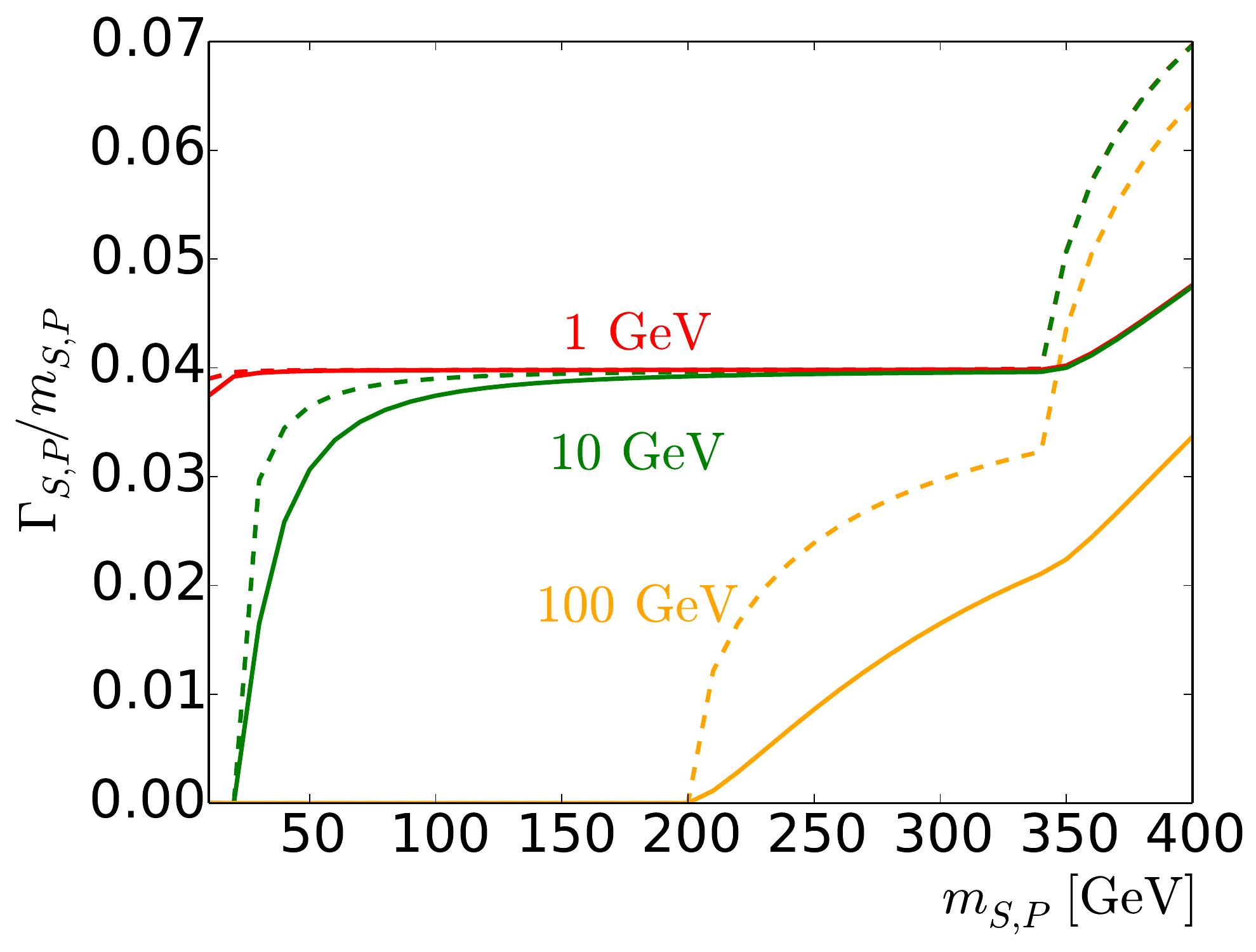}
\caption{Left: production rate for single-top-associated (red) and
  top-pair-associated (blue) dark matter production with an on-shell
  pseudo-scalar (dashed) and scalar (solid) mediator. Right:
  width-to-mass ratios. We assume $g_{S,P}^t=g_{S,P}^\chi=1$ and a
  default mass value of $m_\chi = 1\,\text{GeV}$.}
\label{fig:rate_p}
\end{figure}
%-----------------------------------

For a pseudo-scalar mediator we can perform exactly the same analysis
as for the scalar mediator. However, the signal rate and kinematic
features will be different.  In Fig.~\ref{fig:rate_p}, we show the
total cross sections and normalized widths for different pseudo-scalar
mediator masses (dashed curves), in a similar way as in
Fig.~\ref{fig:rate_s}. The values for a scalar mediator are shown as
plain curves for comparison. The decay width of a pseudo-scalar
mediator is given by
\begin{align}
\frac{\Gamma_P}{m_P}
&= \frac{1}{8\pi}\Bigg[(g_P^\chi)^2 \left(1-\frac{4m_\chi^2}{m_P^2}\right)^{1/2} 
   + 3 (g_P^t)^2 \frac{m_t^2}{v^2}\left(1-\frac{4m_t^2}{m_P^2}\right)^{1/2} \Theta(m_P - 2m_t) \Bigg] \; .
\end{align}
Comparing with Eq.~\eqref{eq:scalar_w}, the scalar width has an
additional threshold suppression of $(1-4m_{\chi,t}^2/m_S^2)$, so that
the scalar decay into fermion pairs proceeds in a $P$-wave~\cite{Han:2014nja,giacomo}. This suppression is absent for the
pseudo-scalar decay, which proceeds in an $S$-wave. The effect on
the pseudo-scalar width is an abrupt increase at $m_P = 2 m_\chi$ and
$m_P= 2 m_t$ observed in the right panel of Fig.~\ref{fig:rate_p}, to
be compared with the smoother on-set of the decay in the scalar
case. Therefore the pseudo-scalar width is larger than the
scalar width for a fixed mediator mass.

The properties of the pseudo-scalar width imprint themselves on the
dark matter production cross section, shown in the left panel of
Fig.~\ref{fig:rate_p}. The kink at the top-pair threshold, $m_P = 2
m_t$, is a consequence of the opening decay channel $P \to t\bar{t}$,
which drastically reduces the branching ratio into dark matter. The
scalar cross section exhibits a smooth top threshold and hence does
not feature this drop. The suppression of the scalar-mediated rate $t\bar t S$ against $t\bar t P$ for
$m_{S,P} \gtrsim 220~\gev$ is also due to the threshold suppression of the process $t\bar t\to S$, which affects
the production rate in this mass region. In the pseudo-scalar case, the
total cross sections for single-top-associated and top-pair-associated
dark matter production are much flatter functions of the mediator
mass. Moreover, for the pseudo-scalar mediator the difference in $tjP$ and $t\bar t P$ cross sections is larger than in the scalar scenario.

%-----------------------------------
\begin{figure}[t]
\includegraphics[width=0.45\hsize]{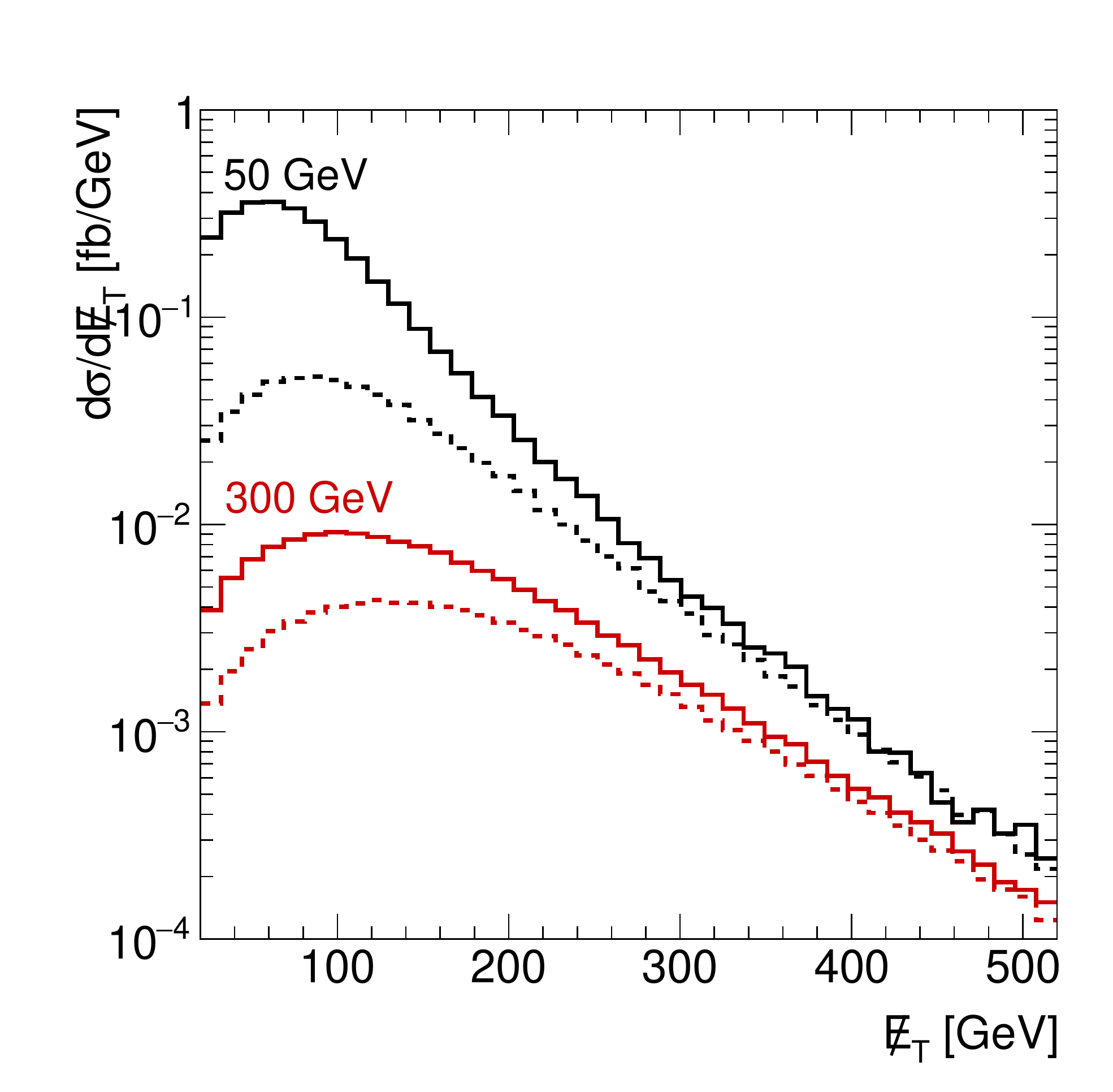}
\hspace*{0.05\hsize}
\includegraphics[width=0.45\hsize]{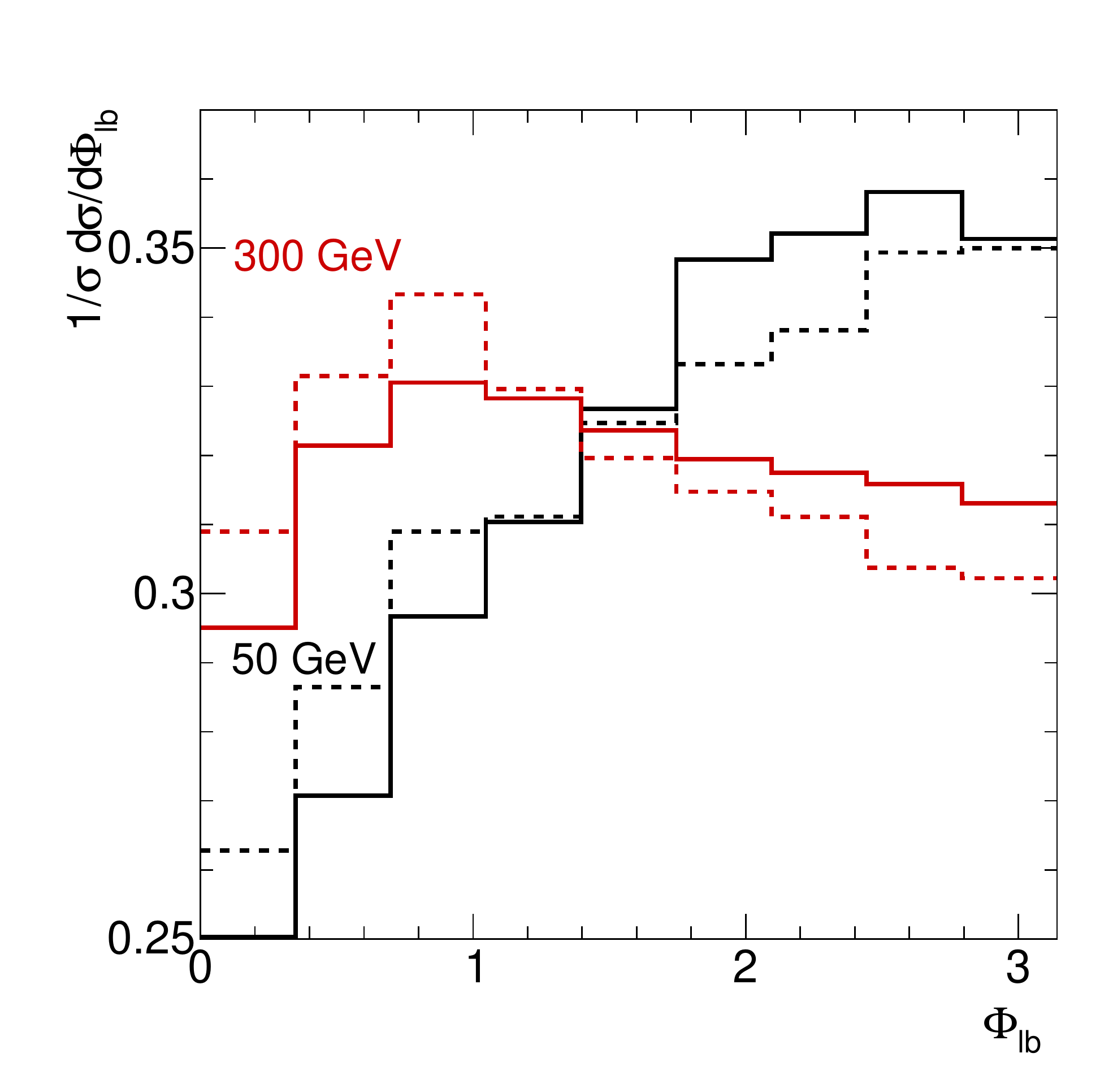}
\caption{Scalar (solid) and pseudo-scalar (dashed) signal distributions of the missing
  transverse energy $\met$ (left) and the azimuthal separation
  $\phi_{\ell,b}$ between the lepton and the $b$-jet (right).  We assume
  mediator masses of $m_{S,P} = 300\,\text{GeV}$ (red) and $m_{S,P} =
  50\,\text{GeV}$ (black), setting the other parameters to $m_\chi =
  1\,\text{GeV}$ and $g_{P,S}^t=g_{P,S}^\chi = 1$.
\label{fig:distri_p1}}
\end{figure}
%-----------------------------------

Extending our comparison to kinematic distributions, we find that for the pseudo-scalar mediator
in single-top-associated dark matter production the $\met$ and $p_T^t$
spectra are harder than for the scalar mediator at the same mass. In
addition, we find a smaller geometric separation between the top and the light-quark jet in the pseudo-scalar case. At large mediator
masses, the production cross sections for scalar- and pseudo-scalar-mediated processes look more similar.

To understand these features, we study the kinematics of the processes
$pp\to t j \chi \bar{\chi}$ and $pp\to t\bar t\chi\bar\chi$. In the
high-energy limit $E_t \gg m_t \gg m_{S,P}$ the radiation of the
mediator off the top quark can be described by the splitting
functions~\cite{effective_higgs,Backovic:2015soa,giacomo}
\begin{align}
f_{t\to tS}(x) & = \frac{(g_S^t)^2}{16\pi^2}\left( x\log \frac{E_t^2}{m_t^2} + 4 \; \frac{1-x}{x} \right), \notag \\
f_{t\to tP}(x) & = \frac{(g_P^t)^2}{16\pi^2} \ x\log \frac{E_t^2}{m_t^2}  \; ,
\end{align}
where $E_t$ is the energy of the incoming top, $E_{S,P}$ the energy of
the mediator, and $x=E_{S,P}/E_t$. The soft enhancement for $x \ll 1$
leads to an enhanced total cross section for light mediators and a
softer $\met$ spectrum as compared to pseudo-scalar mediators. Since
the $\met$ spectrum for the top pair background is softer, the absence
of the soft enhancement in the pseudo-scalar case should make it
easier to extract the corresponding dark matter signal and partly
compensate for the lower production rate.  In the left panel of Fig.~\ref{fig:distri_p1}, 
we show the distributions in missing energy for scalar (plain) and
pseudo-scalar (dashed) mediators. Indeed, the missing energy spectrum for the
pseudo-scalar mediator is harder, missing the soft enhancement. For the kinematics of dark matter associated $t$-channel single top production, shown in Fig.~\ref{fig:feyn}, the
soft enhancement of the scalar emission causes the top and the
light-flavor jet to recoil against each other, while the scalar
mediator does not carry a lot of transverse momentum. In the absence
of the soft enhancement, a pseudo-scalar mediator will balance with
the momenta of the top and the light-flavor jet, pushing them
geometrically closer. Furthermore, in the pseudo-scalar scenario the azimuthal angular separation between the lepton and $b$-jet from the top decay is smaller than for the scalar, as can be seen in the right panel of Fig.~\ref{fig:distri_p1}. Besides being powerful variables to
suppress the background, such kinematic features can also discriminate
between scalar and pseudo-scalar mediators in the case of a discovery.

Using the same setup as in Sec.~\ref{sec:multivariate}, we estimate
the LHC reach for a pseudo-scalar mediator. Again, the set of
kinematic variables shown in Eq.~\eqref{eq:multivariate} separate the
signal and the backgrounds in a multi-variate BDT analysis combined
with a $\text{CL}_S$ analysis based on a variable cut on the BDT
output variable. As discussed above, this setup avoids large
systematic uncertainties from counting events in the tails of
distributions. In the right panel of Fig.~\ref{fig:significance_s}, we show the signal
strength for a pseudo-scalar mediator that can be excluded at the
$95\%$ CL by the LHC. Compared with the scalar, the sensitivity is
lower throughout the considered range of mediator masses. This is
mainly due to the lower production rate of the pseudo-scalar
mediator. For $m_P = 50\,\text{GeV}$, couplings
$g_P^t > 0.8 (0.4)$ can be excluded with $300\,\text{fb}^{-1}$
($3\,\text{ab}^{-1}$) of data. In the intermediate mass range around $m_P = 200~\gev$, the sensitivity to pseudo-scalar mediators is somewhat higher than to scalars. At high mediator masses, a signal strength of $\mu=1$ can ultimately be excluded for $m_P < 340~\gev$. Compared with top-pair-associated production, the sensitivity to pseudo-scalars is slightly lower in the considered region of mediator masses.

%%%%%%%%%%%%%%%%%%%%%%%%%%%%%%%%%%%%%%%%%%%%%%%%%%%%%%%%%%%%%%%%%%%%%%
\section{Summary}
Fermion dark matter interacting dominantly with top quarks through a new scalar mediator can be probed at the LHC through associated production with tops. While searches for missing energy in association with a top-anti-top pair are well established, dark matter production in association with a single top quark has started to be considered only recently. In this work, we propose $t$-channel single top production with large missing transverse energy as a new probe of invisible particles with large top-quark couplings. This signature can be efficiently separated from Standard Model backgrounds, as well as from
dark matter production with a top-anti-top pair, by exploiting the kinematic features of electroweak single top production. In particular, the hard forward jet is an important characteristic of the signal and clearly distinguishes it from other searches, such as mono-tops.

Our signal employs leptonically decaying top quarks, requesting one lepton, one $b$-jet, large missing energy, as well as one hard forward jet in the event selection. The dominant background in this phase-space region is due to top-anti-top production with leptonic decays of both tops. In a dedicated analysis based on boosted decision trees, we take full account of the kinematic differences between signal and background. Powerful discriminators are transverse mass variables that strongly suppress topologies with additional $W$ bosons or top quarks, as well as large missing energy and the forward jet kinematics. Since our analysis is statistics limited, an optimal signal sensitivity involves phase-space regions with significant top-anti-top background. In practice, this well-known background should be measured in control regions to limit systematical uncertainties.

At the LHC, we expect that single-top-associated dark matter production can be tested for on-shell scalar mediators up to masses of $m_S = 180\,(360)\,\text{GeV}$ with $300\,\text{fb}^{-1}(3\,\text{ab}^{-1})$ of data, assuming a coupling strength of $g_S^t=1$. For pseudo-scalar mediators, the mass reach extends to $m_P = 230\,(340)\,\text{GeV}$. Compared with top pair associated dark matter
production, the sensitivity to single-top-associated production is similar for low mediator masses and somewhat smaller for
high masses. The prospects to find invisible particles with top couplings at the LHC can thus be clearly
improved by performing a dedicated search for single-top-associated dark matter production. As the theory hypothesis
behind both signals is exactly the same, the two searches should be combined to maximize the sensitivity. In case of an observation, the combination of $tj\chi\bar{\chi}$ and $t\bar t\chi\bar{\chi}$ analyses can be used to understand the properties of the dark sector. In particular, the parity quantum number of the mediator can be determined from the different kinematics of scalar and pseudo-scalar, which originates from their threshold behavior in scattering processes.

\acknowledgments We acknowledge support by the DFG research unit
\textit{New Physics at the LHC} (FOR2239).  SW acknowledges support
by the Carl Zeiss Foundation through an endowed junior professorship.

\newpage

%%%%%%%%%%%%%%%%%%%%%%%%%%%%%%%%%%%%%%%%%%%%%%%%%%%%%%%%%%%%%%%%%%%%%%

\end{fmffile}
\end{document}